\documentclass{mem}
\usepackage{natbib}
\usepackage{txfonts}
\usepackage{balance}
\usepackage{graphicx}

\usepackage{bm}
\usepackage{amssymb}
\usepackage{enumerate}

\def\ba{{\bm a}}
\def\bb{{\bm b}}

\def\bN{{\bm N}}

\def\bx{{\bm x}}

\def\lb{\label}

\def\be{\begin{equation}}
\def\ee{\end{equation}}
\def\bea{\begin{eqnarray}}
\def\eea{\end{eqnarray}}
\idline{75}{282}
\begin{document}
\def\teff{$T\rm_{eff }$}
\def\kms{$\mathrm {km s}^{-1}$}

\title{
Light deflection for relativistic space astrometry in closed form
}

   \subtitle{}

\author{
Stefano Bertone\inst{1} 
\and Christophe Le Poncin-Lafitte
          }

  \offprints{\\stefano.bertone@obspm.fr}

\institute{
Observatoire de Paris - SYRTE,
CNRS/UMR 8630, Universit\'e Pierre et Marie Curie - 
77, Avenue Denfert Rochereau, F-75014 Paris, France
}

\authorrunning{Bertone}

\titlerunning{Time transfer functions for relativistic space astrometry}

\abstract{
Given the extreme accuracy reached by future global space astrometry, one needs a global relativistic modeling of observations. A relativistic definition of astrometric observables is then essential to find uniquely coordinates, parallax and proper motion of a stellar object. The standard procedure is to explicitly solve the null geodesic equations to describe the trajectory of a photon emitted by a celestial object and received by a moving observing satellite. However, this task can be avoided if one builds an astrometric set up using the recent formalism of time transfer functions.
}
\maketitle{}

%%%%%%%%%%%%%%%%%%%
\section{Introduction \label{sect:intro}}
%%%%%%%%%%%%%%%%%%%
\par Future space astrometry missions, such as Gaia~\citep{2002EAS.....2.....B}, will soon provide large astrometric catalogues with typical accuracy of several microarcseconds ($\mu$as). It is nowadays well-known that $\mu$as astrometry needs a precise relativistic modeling to define and calculate accurate astrometric parameters. In particular, when a satellite observes a celestial object, one needs to know its attitude and onboard relativistic proper time as well as the deflection of the incoming light ray detected on the CCD. In the context of the Gaia mission, several approaches are available, notably GREM~\citep{2004PhRvD..69l4001K} and RAMOD~\citep{2006ApJ...653.1552D}. Both modelings need to determine the full light trajectory from the emitting celestial object to the observing satellite by solving the null geodesic equations. However, it has been recently demonstrated that this task is not at all mandatory and can be replaced by another approach, initially based on the Synge World Function \citep{2004CQGra..21.4463L} and then on the time transfer functions \citep{2008CQGra..25n5020T}. 
The purpose of this article is to build straightforwardly an astrometric set up using the formalism of the time transfer functions.

\par This paper is organized as follows. Section~\ref{sect:2} gives the notations used in this article. 
In section~\ref{sect:3} we set up the astrometric problem. 
Section~\ref{sect:4} recalls the relation between the covariant components of the vector tangent to a light signal and its flight time between two events. 
This relation is then explicitly given in section~\ref{sect:5} within the post-Newtonian approximation.
Finally, section~\ref{sect:concl} delivers some concluding remarks.

 %%%%%%%%%%%%%%%%%%%%%%%%%%%%%%%%%%
\section{Notations and conventions \label{sect:2}}  %%
%%%%%%%%%%%%%%%%%%%%%%%%%%%%%%%%%%
\par Throughout this work, $c$ is the speed of light in vacuum and $G$ is the Newtonian gravitational constant. The Lorentzian metric of space-time $V_4$ is denoted by $g$. We adopt the signature $(+ - - -)$. We suppose that space-time is covered by some global coordinate system $x^{\alpha} = (x^0, \bx)$, with $x^0 = ct$ and $\bx = (x^i)$, centered on the Solar System barycenter. Greek indices run from 0 to 3 and Latin indices from 1 to 3. Moreover, we assume that the curves of equation $x^i =$ const are time-like, at least in the neighborhood of the chosen observer. This condition means that $g_{00}> 0$ in the vicinity of the observer. Any ordered triple is denoted by a bold letter. In order to distinguish the triples built with the space-like contravariant components of a vector from the ones built with covariant components, we systematically use the notation ${\ba} =(a^1,a^2,a^3)=(a^i)$ and $\underline{\bb}= (b_1,b_2,b_3)=(b_i)$. 

 %%%%%%%%%%%%%%%%%%%%%%%%%%%%%%%%%%
\section{Set up of the astrometric problem \label{sect:3}}  %%
%%%%%%%%%%%%%%%%%%%%%%%%%%%%%%%%%%
\par Let us consider a time-like observer. All along its worldline, we choose a tetrad of four vectors $E_{(\alpha)}^\mu$, where index $(\alpha)$ is only the tetrad index, which runs from 0 to 3 to enumerate the 4-vectors, while $\mu$ is a normal tensor index which can be lowered and raised by the use of the metric tensor :
\begin{equation}
E_{(\alpha)\mu}=g_{\mu\nu}\, E_{(\alpha)}^\nu \,,\quad E_{(\alpha)}^{\mu}=g^{\mu\nu}E_{(\alpha) \, \nu}\,.
\end{equation}  
The tetrad vectors are required to satisfy the following property
\begin{equation}
E_{(\alpha)}^{\mu}E_{(\beta)\, \mu}=\eta_{\alpha \beta}\, ,
\end{equation}
which implies that the vectors are orthogonal to each other, vector $E_{(0)}^\mu$ being unit and time-like, and $E_{(i)}^\mu$ being unit and space-like. In the following, we do not specify more additional information about this tetrad.

Let us now study the reception of a light ray from a celestial object by a satellite. Its direction is given by the vector $k^\mu=(k^0, k^i)$, tangent to the incoming light ray and its unit space-like direction $\tilde{k}^\mu$ relative to the hyper-surface orthogonal to $E_{(0)}^\mu$ is given by \cite{2006gr.qc....11078T}
\begin{equation}
\tilde{k}^\mu=\frac{k^\mu}{E_{(0)}^\nu k_\nu}-E_{(0)}^\mu\, \label{eqt: ktilde}.
\end{equation}
Then, we calculate the three director cosines $\cos \phi_{(a)}$ formed by each space-like vector $E_{(a)}^\nu$ (here index $(a)$ runs from 1 to 3) with $\tilde{k}^\mu$
\begin{equation}
\cos \phi_{(a)}= - g_{\mu\nu} \, \tilde{k}^\mu\, E_{(a)}^{\nu}\, \label{eqt: cosphi}.
\end{equation}
Taking Eqs.~(\ref{eqt: ktilde}~-~\ref{eqt: cosphi}) into account, the final expression of $\cos \phi_{(a)}$ can be written as follows
\begin{equation}
\cos \phi_{(a)}=-\frac{E_{(a)}^0+\hat{k}_i \, E_{(a)}^i}{E_{(0)}^0\left(1+\hat{k}_i \, \beta^i\right)} \, , \label{eqt:dircos}
\end{equation}
where $\beta^i=v^i/c$, $v^i$ is the coordinate velocity of the observer and $\hat{k}_i=k_i/k_0$ will be called in the following the {\em deflection functions}. 

\par Eq.~(\ref{eqt:dircos}) shows that we only need the deflection functions at the reception event to determine the director cosines of an astrometric observation and that it's not necessary to trace the light ray from the emission event to the reception event. In the next session, we recall the relation between time delay and light deflection.
  
%%%%%%%%%%%%%%%%%%%%%%%%%%%%%%%%%%%%%%%%%%%%
\section{From coordinate time delay to light deflection \label{sect:4}} 
%%%%%%%%%%%%%%%%%%%%%%%%%%%%%%%%%%%%%%%%%%%%

\par Let us consider that space-time is globally regular with the topology $\Bbb{R}\times \Bbb{R}^3$ and that it is without horizon, which is admissible in the context of practical space astrometry within the Solar System. We suppose the existence of a unique light ray connecting $x_A = (ct_A, \bx_A)$, the event of emission of a photon from a celestial object, and $x_B = (ct_B, {\bm x}_B)$, the event of reception of the light by the observing satellite. The difference $t_B - t_A$ is the (coordinate) travel time of the photon between the two points. This quantity may be considered either as a function of the instant of reception $t_B$ and of  ${\bm x}_A$, ${\bm x}_B$, or as a function of the instant of emission $t_A$ and of ${\bm x}_A$ and ${\bm x}_B$. So, it is possible to define two time transfer functions ${\mathcal T}_r$ and ${\mathcal T}_e$ by putting\footnote{Note that the order of the arguments of ${\cal T}_{r}$ in Eq.~(\ref{ttf}) is different from the one given in Ref.~\citep{2004CQGra..21.4463L}. 
} 
\begin{equation} \label{ttf}
t_B - t_A = {\mathcal T}_r({\bm x}_A, t_B, {\bm x}_B) = {\mathcal T}_e(t_A, {\bm x}_A, {\bm x}_B),
\end{equation}

where ${\cal T}_r$ and ${\cal T}_e$ are called the reception and the emission time transfer function, respectively. Explicit expressions of these functions are different except in a stationary space-time in which the coordinate system is chosen so that the metric does not depend on $x^0$. Here, we are obviously most interested in ${\cal T}_r$. Indeed, when the satellite is doing an observation, we know its state's vector and its onboard time.

\par Moreover, a theorem relating the deflection functions and ${\cal T}_r$ has been found by~\citep{2004CQGra..21.4463L}, where $\hat{k}_i$ is obtained from the partial derivatives of ${\cal T}_r$ as follow
\be \lb{eq:kik0}
\left( \hat{k}_i \right)_{x_B} =- c \, \frac{\partial  {\cal T}_{r}}{\partial x^{i}_{B}}
\left[1 - \frac{\partial  {\cal T}_{r}} {\partial t_B}\right]^{-1}\,  .
\ee

The question is then how to determine this function in order to calculate the astrometric director cosines. 

%%%%%%%%%%%%%%%%%%%%%%%%%%%%%%%%%%%%%%%%%%%%%%%%%%%
\section{Post-Newtonian time transfer and deflections functions \label{sect:5}}   %%
%%%%%%%%%%%%%%%%%%%%%%%%%%%%%%%%%%%%%%%%%%%%%%%%%%%

It was shown that ${\cal T}_{r}$ is the solution of the following partial differential equation~\citep{2008CQGra..25n5020T}
\bea 
&& \!\!\!\! g^{00}(x^0_B - c{\cal T}_{r},\bx_A)+ 2c \, g^{0i}(x^0_B - c{\cal T}_{r},\bx_A)\frac{\partial {\cal T}_{r}}{\partial x^{i}_{A}} \nonumber \\
&&+ c^2\,g^{ij}(x^0_B-c{\cal T}_{r},\bx_A)\frac{\partial {\cal T}_{r}}{\partial x^{i}_{A}}\frac{\partial {\cal
T}_{r}}{\partial x^{j}_{A}} = 0 \; . \label{eiko}
\eea
\par Finding a solution to this equation is as challenging than a determination of the null geodesic equations~\citep{1999PhRvD..60l4002K}. However, in the weak field approximation, this task is easier. Let us write the metric tensor as follow
\begin{equation}
g_{\mu\nu}=\eta_{\mu\nu}+h_{\mu\nu}\, ,
\end{equation}
with $\eta$ a Minkowskian background and $h$ a perturbation admitting the following general post-Minkowskian expansion
\begin{equation}
h_{\mu\nu}=\sum_{n=1}^\infty G^n h_{\mu\nu}^{(n)}\, .
\end{equation}
It was possible to show that ${\cal T}_{r}$ can be written as a series of ascending powers of the Newtonian gravitational constant $G$
\begin{equation}
{\cal T}_{r}(t_B, \bx_A, \bx_B)=\frac{R}{c}+\sum_{n=1}^\infty G^n {\cal T}_{r}^{(n)}(t_B, \bx_A, \bx_B)\, ,
\end{equation}
where it has been demonstrated that each perturbation term $\mathcal{T}_r^{(n)}$ can be obtained as a line integral along a null straight line. This result is particularly interesting and can be interpreted as a Fermat principle~\citep{1990CQGra...7.1319P} in the $n$th-post-Minkowskian approximation. 

In particular, if we work in the context of the post-Newtonian approximation, and taking the metric tensor recommended by the International Astronomical Union in 2000~\citep{2003AJ....126.2687S}, we can write the metric tensor $g_{\mu\nu}$ for a light signal as
\begin{eqnarray} \lb{1n}
&& g_{00}=1-\frac{2}{c^2}W(t, {\bm x})+\mathcal{O}\left(c^{-4}\right) \nonumber \, , \\
&& g_{0i}=\frac{2\,(1+\gamma)}{c^3}W^i(t, {\bm x})+\mathcal{O}\left(c^{-5}\right) \, , \\ 
&& g_{ij}=- \left( 1+\frac{2}{c^2}\, \gamma \,W(t, {\bm x}) \right) \delta_{ij} +\mathcal{O}\left(c^{-4}\right) \, , \nonumber
\end{eqnarray}
where $\gamma$ is a PPN parameter~\citep{1993tegp.book.....W}, $W(t,\bx)$ and ${\bm W}=\lbrace W^i(t,\bx)\rbrace$ being the gravitational potentials at the coordinate $(t,\bx)$.
After substitution of the metric into Eq.~(\ref{eiko}), a straightforward calculation leads to 
\begin{eqnarray}  \label{eq:Tr}
&&\!\!\!\!\! {\cal T}_{r}(t_{B}, {\bm x}_{A}, {\bm x}_{B}) = \frac{R}{c}\\
&&\!\!\!\!\! +\frac{R}{c^3}\int_{0}^{1}\left[ (\gamma+1)\, W -\frac{2\,(1+\gamma)}{c} \bm W \! \cdot \! \bN \right]_{z_{-}(\lambda)} \!\!\! d\lambda+\mathcal{O}\left(c^{-5}\right), \nonumber
\end{eqnarray} 
where the integral is taken along the null straight line
\begin{equation} \label{13r}
z_{-}(\lambda)=(ct_B-\lambda R, {\bm x}_B-\lambda R{\bm N})\, ,
\end{equation}
with $R=\vert{\bm x}_B-{\bm x}_A \vert$ and ${\bm N}=\left\{N^i\right\}=({\bm x}_B-{\bm x}_A)/R$ being the Minkowskian direction between $\bx_A$ and $\bx_B$.

Considering Eqs.~(\ref{eq:Tr}~-~\ref{13r}), it is possible to calculate the partial derivatives of $\mathcal{T}_r$. One gets 
\begin{eqnarray} \lb{eq:dtTr}
&&\!\!\!\!\! \partial_{t_B}{\mathcal T}_r =\frac{(\gamma+1) \, R}{c^3}\int_0^1\left[\frac{\partial W}{\partial t}\right] _{z_{-}^\alpha}d\lambda \\
&&\quad -\frac{2\,(1+\gamma) R}{c^4}\int_0^1\left[\left({\bm N} \cdot \frac{\partial{\bm W}}{\partial t} \right)\right] _{z_{-}^\alpha}d\lambda+\mathcal{O}\left(c^{-5}\right)\, ,   \nonumber 
\end{eqnarray}
and
\bea  \lb{eq:dxTr}
&&\!\!\!\!\! \partial_{x_B^i}{\mathcal T}_r = \frac{N^i}{c} \\
&&\quad +\frac{(\gamma+1)}{c^3}\int_0^1\left[W\,N^i+(1-\lambda) \, R\frac{\partial W}{\partial x^i}\right]_{z_{-}^\alpha} d\lambda \nonumber\\
&&\quad -\frac{2\,(1+\gamma)}{c^4}\int_0^1\left[ W^i+\frac{1}{2} \lambda \, R \, N^i \frac{\partial W}{\partial t}\right. \nonumber \\
&&\quad \left.+(1-\lambda) \, R \left({\bm N} \cdot \frac{\partial{\bm W}}{\partial x^i} \right)  \right]_{z_{-}^\alpha} d\lambda+\mathcal{O}\left(c^{-5}\right) . \nonumber
\eea
Substituting Eqs.~(\ref{eq:dtTr}~-~\ref{eq:dxTr}) into Eq.~(\ref{eq:kik0}), we obtain finally the following relation for the deflection functions
\bea
&&\!\!\!\!\! \left( \hat{k}_i \right)_{x_B}=-N^i \\
&&\quad - \frac{(\gamma +1)}{c^2} \int_0^1\left[W \, N^i+(1-\lambda) \, R \frac{\partial W}{\partial x^i} \right]_{z_{-}^\alpha(\lambda)}d\lambda  \nonumber\\
&&\quad +\frac{2\,(1+\gamma)}{c^3} \int_0^1\left[ W^i-\frac{1}{2} \, (1-\lambda) \, R \, N^i \frac{\partial W}{\partial t}\right.\nonumber\\
&&\qquad \left.+(1-\lambda) \, R \left( {\bm N}\,.\frac{\partial {\bm W}}{\partial x^i} \right) \right]_{z_{-}^\alpha(\lambda)}d\lambda +\mathcal{O}\left(c^{-4}\right)\nonumber \label{deflexionGen} .
\eea

%%%%%%%%%%%%%%%%%%%%%%
\section{Conclusions \label{sect:concl}}
%%%%%%%%%%%%%%%%%%%%%%%

\par We presented in this article a relativistic procedure for global astrometry based on the time transfer functions formalism. We were able to obtain the light deflection of a light ray connecting a stellar source and an observing satellite solving the boundary value problem. The deflection functions in the Post-Newtonian approximation were then obtained as a line integral of the gravitational potentials along the Minkowskian line of sight. This approach is quite new and needs further development but, in principle, it avoids the difficult task of solving the null geodesic equations.

\bibliographystyle{aa}  % A&A bibliography style file (aa.bst)
\bibliography{/Users/stefano/Documents/biblio/BiblioPerso} % your references in file: Yourfile.bib

\begin{thebibliography}{10}
\expandafter\ifx\csname natexlab\endcsname\relax\def\natexlab#1{#1}\fi

\bibitem[{Bienayme \& Turon(2002)}]{2002EAS.....2.....B}
Bienayme, O. \& Turon, C. 2002, EAS Publications Series, 2

\bibitem[{de~Felice {et~al.}(2006)de~Felice, Vecchiato, Crosta, Bucciarelli, \&
  Lattanzi}]{2006ApJ...653.1552D}
de~Felice, F., Vecchiato, A., Crosta, M.~T., Bucciarelli, B., \& Lattanzi,
  M.~G. 2006, Astrophysical Journal, 653, 1552

\bibitem[{Klioner(2004)}]{2004PhRvD..69l4001K}
Klioner, S.~A. 2004, Physical Review D, 69, 124001

\bibitem[{Kopeikin \& Sch{\"a}fer(1999)}]{1999PhRvD..60l4002K}
Kopeikin, S.~M. \& Sch{\"a}fer, G. 1999, Physical Review D, 60, 124002

\bibitem[{Le~Poncin-Lafitte {et~al.}(2004)Le~Poncin-Lafitte, Linet, \&
  Teyssandier}]{2004CQGra..21.4463L}
Le~Poncin-Lafitte, C., Linet, B., \& Teyssandier, P. 2004, Classical and
  Quantum Gravity, 21, 4463

\bibitem[{Perlick(1990)}]{1990CQGra...7.1319P}
Perlick, V. 1990, Classical and Quantum Gravity, 7, 1319

\bibitem[{Soffel {et~al.}(2003)Soffel, Klioner, Petit, Wolf, Kopeikin,
  Bretagnon, Brumberg, Capitaine, Damour, Fukushima, Guinot, Huang, Lindegren,
  Ma, Nordtvedt, Ries, Seidelmann, Vokrouhlick{\'y}, Will, \&
  Xu}]{2003AJ....126.2687S}
Soffel, M., Klioner, S.~A., Petit, G., {et~al.} 2003, Astronomical Journal,
  126, 2687

\bibitem[{Teyssandier \& Le~Poncin-Lafitte(2006)}]{2006gr.qc....11078T}
Teyssandier, P. \& Le~Poncin-Lafitte, C. 2006, ArXiv General Relativity and
  Quantum Cosmology e-prints

\bibitem[{Teyssandier \& Le~Poncin-Lafitte(2008)}]{2008CQGra..25n5020T}
Teyssandier, P. \& Le~Poncin-Lafitte, C. 2008, Classical and Quantum Gravity,
  25, 145020

\bibitem[{Will(1993)}]{1993tegp.book.....W}
Will, C.~M. 1993, Theory and Experiment in Gravitational Physics, ed. C.~M.
  Will

\end{thebibliography}

\end{document}